\documentclass[twocolumn,english,aps,prl,showpacs]{revtex4}
\usepackage[T1]{fontenc}
\usepackage{amsmath}
\usepackage{graphicx}
\usepackage{amssymb}
\usepackage{esint}

\usepackage{babel}
\usepackage{dsfont}

\renewcommand{\Re}{{\rm Re}}
\renewcommand{\Im}{{\rm Im}}
\newcommand{\Tr}{{\rm Tr}}
\newcommand{\rd}{{\rm d}}
\newcommand{\kb}{k_{\rm B}}

\usepackage{babel}

\begin{document}

\title{Many-body radiative heat transfer theory}

\author{Philippe Ben-Abdallah}

\author{Svend-Age Biehs}
\altaffiliation[Present address: ]{Institut f\"{u}r Physik, Carl von Ossietzky Universit\"{a}t,
D-26111 Oldenburg, Germany.}
        
\affiliation{Laboratoire Charles Fabry,UMR 8501, Institut d'Optique, CNRS, Universit\'{e} Paris-Sud 11,
2, Avenue Augustin Fresnel, 91127 Palaiseau Cedex, France.}

\author{Karl Joulain}

\affiliation{Institut P', CNRS-Universit\'{e} de Poitiers-CNRS UPR 3346, 86022 Poitiers
Cedex, France.}

\date{\today}


\begin{abstract}
In this Letter a $N$-body theory for the radiative heat exchange
in thermally non equilibrated discrete systems of finite size objects
is presented. We report strong exaltation effects of heat flux which
can be explained only by taking into account the presence of many
body interactions. Our theory extends the standard Polder and van
Hove stochastic formalism used to evaluate heat exchanges between
two objects isolated from their environment to a collection of objects
in mutual interaction. It gives a natural theoretical framework to
investigate the photon heat transport properties of complex 
systems at mesoscopic scale. 
\end{abstract}

\maketitle

The photon heat tunneling between two bodies has attracted much attention
in the last decades since it has been predicted that the heat flux (HF)
can exceed, at nanoscale, the far field limit set by Planck's black
body law by several orders of magnitude~\cite{Polder1973,Planck}.
This discovery has opened the way to promising technologies for energy
conversion and data storage as for example the near-field 
thermophotovoltaics~\cite{MatteoEtAl2001,NarayanaswamyChen2003} and
the plasmon assisted nanophotolitography~\cite{Srituravanich}. This
dramatic increase is generally speaking due to the contribution
of evanescent modes, which are not accounted for in the Stefan-Boltzmann
law and become only important if the distance between the objects
is smaller than the thermal wavelength~\cite{VolokitinPersson07}.
The detailed mechanisms which lead to such an enhancement are nowadays
for a number of geometries and materials well understood~\cite{VolokitinPersson07,Nara2008,Domingues2005,Perez2008,ChapuisEtAl2008,MuletEtAl2001,Dorofeyev1999,Biehs2007,PBA2009,PBA2010,Joulain2010,Rueting2010,BiehsEtAl2011,BiehsEtAl2011b,Krueger2011,Zhang2005,RousseauEtAl2009b}
and recent experiments~\cite{Kittel,HuEtAl2008,ShenEtAl2008,RousseauEtAl2009,Ottens2011}
have confirmed all theoretical predictions both qualitatively and
quantitatively.

However, some questions of fundamental importance remain unsolved
in complex mesoscopic systems. Indeed, so far, only the HF 
between two objects~\cite{Nara2008,Domingues2005,Perez2008,VolokitinPersson07}
out of equillibrium has been considered, but how does the heat transport
for a collection of individual objects in mutual interaction look
like? The collective effects in such many particle systems has not
been explored yet, although it is of prime importance for understanding
the different heat propagation regimes in disordered systems, determining
the thermal percolation tresholds in random nanocomposites structures
and studying  thermal effects due to the presence of localized modes in such systems. 

Inside a discrete system of bodies maintained at different temperatures
the local thermal fluctuations give rise to oscillations of partial
charges which, in turn, radiate their own time dependent electric
field in the surrounding medium. These thermally generated fields
interact with the nearby bodies and modify through different cross
interactions all these primary fields to generate secondary fields
which in turn affect the radiated fields and so on. Generally speaking,
this problem belongs to the vast category of many-body problems which
constitute the theoretical framework of numerous branches of physics
(celestial mechanics,condensed matter physics, atomic physics, quantum chemistry). 
A general theoretical framework
to treat the many-body problem of non-radiative photon heat transport
does not yet exist. In this Letter we introduce a self-consistent
theory to describe heat transfers inside thermally non equilibrated
discrete systems. After deriving the HF exchanged between two
individual objects in mutual interaction inside a N-body system, we
investigate the thermal conductance between a couple of particles
versus the position of a third object inside a three body system.
We highlight some emergent phenomena which specifically result from
many-body interactions. In addition, we will show that for systems with 
at least three objects at different temperatures one  
can actively control the heat flow in nanoscale junctions, i.e., one
has a thermal heat transfer transistor.

To start, let us consider a discrete set of $N$ objects located at
positions $r_{i}$ and maintained at different temperatures $T_{i}$
with $i=1,\ldots,N$. Suppose that the size of these objects is small
enough compared with the smallest thermal wavelength $\lambda_{T_{i}}=c\hbar/(\kb T_{i})$
so that all individual objects can be modeled to simple radiating electrical
dipoles. For metals one has also to include the magnetic dipole moments due to
the induction of eddy currents~\cite{ChapuisEtAl2008}. Such an extension of our approach is straightforward so that
for convenience we will consider electric dipoles only. The Fourier component of the electric field at the frequency
$\omega$ (with the convention $\widehat{f}(t)=\int\frac{d\omega}{2\pi}f(\omega)e^{-i\omega t}$)
generated at the position $r_{i}$ by the fluctuating part  $\mathbf{p}_{j}^{\rm fluc}$ of electric dipole moment 
of the particle $j$ which is located at $r_{j}$ reads 
\begin{equation}
  \mathbf{E}_{ij}=\omega^{2}\mu_{0}\mathds{G}^{ij}\mathbf{p}_{j}^{\rm fluc},
  \label{Eq:GreensFunction}
\end{equation}
 with $\mu_{0}$ the vacuum permeability and $\mathrm{{\normalcolor \mathds{G}^{ij}\equiv\mathds{G}(r_{i},r_{j};\omega)}}$
the dyadic Green tensor (i.e. the propagator) between the particles
$i$ and $j$ inside the set of N particles. On the other hand, by
suming the contribution of fields radiated by each particle, the dipolar
moment induced by the total field on the $i$-th particle is given
by 
\begin{equation}
  \mathbf{p}_{i}^{\rm ind}=\varepsilon_{0}\alpha_{i}\underset{j\neq i}{\sum}\mathbf{E}_{ij}\label{Eq:Polarizability},
\end{equation}
where $\alpha_{i}$ is the particle's polarizability and $\varepsilon_{0}$
is the vacuum permittivity. Then, the power dissipated inside the
particle $i$ at a given frequency $\omega$ by the fluctuating field
$\mathbf{E}_{ij}$ generated by the particle $j$ can be calculated
from the work of the fluctuating electromagnetic field on the charge
carriers as 
\begin{equation}
  \mathcal{P}_{j\rightarrow i,\omega}=2\Re\langle-i\omega\mathbf{p}_{i}^{\rm ind}(\omega)\cdot\mathbf{E}_{ij}^{*}\rangle,
\end{equation}
 where the brackets represent the ensemble average. Using relations
(\ref{Eq:GreensFunction}) and (\ref{Eq:Polarizability}) between
the dipole moments and the fluctuation dissipation theorem, i.e.\
$\langle p_{j,\alpha}^{*}p_{i,\beta}\rangle=2\frac{\epsilon_{0}}{\omega}\Im(\alpha_{j})\Theta(\omega,T_{j})\delta_{\alpha\beta}\delta{ij}$ 
we find after a straightforward calculation that 
\begin{equation}
  \mathcal{P}_{j\rightarrow i}=3\int_{0}^{\infty}\frac{\rd\omega}{2\pi}\,\Theta(\omega,T_{j})\mathcal{T}_{i,j}(\omega)\label{Eq:InterpartHeatFlux}
\end{equation}
 introducing the transmission coefficient (TC)
\begin{equation}
  \mathcal{T}_{i,j}(\omega)=\frac{4}{3}\frac{\omega^{4}}{c^{4}}\Im(\alpha_{i})\Im(\alpha_{j})\Tr\bigl[\mathds{G}^{ij}\mathds{G}^{ij\dagger}\bigr].
\end{equation}
In order to present the HF in an obvious Landauer-like manner~\cite{BiehsEtAl2010,JoulainPBA2010}
we rewrite the HF in terms of the conductance $G_{i,j}=\partial\mathcal{P}_{j\rightarrow i}/\partial T_{j}$
so that $\mathcal{P}_{j\rightarrow i}=G_{i,j}\Delta T$. Then we find
\begin{equation}
  \mathcal{P}_{j\rightarrow i}=3\biggl(\frac{\pi^{2}\kb T}{3h}\biggr)\overline{\mathcal{T}}_{i,j}\Delta T
  \label{Eq:InterpartHeatFluxCond}
\end{equation}
where $\overline{\mathcal{T}}_{i,j}=\int\rd x\, f(x)\mathcal{T}_{i,j}(x)/(\pi^{2}/3)$
is the mean TC~\cite{BiehsEtAl2010} with $f(x)=x^{2}\exp(-x)/(\exp(x)-1)^{2}$.
This expression generalizes the Meir-Wingreen Landauer type formula~\cite{Ojanen} for photon HF 
in N-body systems. In the case of two particles ($N=2$) one
can easily show that $\mathcal{T}_{i,j}(\omega,d)\in[0,1]$ and therefore
$\overline{\mathcal{T}}_{i,j}\in[0,1]$ as well. Hence the conductance between
two dipoles is limited by 3 times the quantum of thermal conductance
$\pi^{2}\kb T/(3h)$~\cite{Pendry1983}. In other words,
only three channels contribute to the HF between two dipoles
namely the channels due to the coupling of the three components $p_{j,\alpha}$
with the same three components $p_{j,\alpha}$ (i.e. same polarization).
Of course, by adding further particles this limit cannot be exceeded,
whereas the HF can be increased or decreased with respect to
the case of two particles. Nevertheless, the number of channels increases
if electric multipoles~\cite{Yanno2007,PBA2008} as well as the magnetic
moments~\cite{ChapuisEtAl2008}
come into play.


Now, for calculating the Green's function (GF) for a system of $N$ particles
we use the set of $3N$ self-consistent equations~\cite{Purcell}
\begin{equation}
\mathbf{E}_{ij}=\mu_{0}\omega^{2}\mathds{G}_{0}^{ij}\mathbf{p}_{j\neq i}^{\rm fluc}+\frac{\omega^{2}}{c^{2}}\underset{k\neq i}{\sum}\mathds{G}_{0}^{ik}\alpha_{k}\mathbf{E}_{kj}\end{equation}
 for $i=1,...,N$ with the free space GF $\mathds{G_{\mathrm{0}}^{\mathrm{ij}}}=\frac{\exp(ikr_{ij})}{4\pi r_{ij}}\left[\left(1+\frac{ikr_{ij}-1}{k^{2}r_{ij}^{2}}\right)\mathds{1}+\frac{3-3ikr_{ij}-k^{2}r_{ij}^{2}}{k^{2}r_{ij}^{2}}\widehat{\mathbf{r}}_{ij}\otimes\widehat{\mathbf{r}}_{ij}\right]$
the vacuum GF defined with the unit vector $\widehat{\mathbf{r}}_{ij}\equiv\mathbf{r_{\mathit{ij}}}/r_{ij}$,
$\mathbf{r_{\mathit{ij}}}$ being the vector linking the center of
dipoles i and j, while $r_{ij}=\mid\mathbf{r}_{ij}\mid$and $\mathds{1}$
stands for the unit dyadic tensor. Inserting relation (\ref{Eq:GreensFunction})
into this sytem leads to the GF \begin{equation}
\left(\begin{array}{c}
\mathds{G}^{1k}\\
\vdots\\
\mathds{G}^{Nk}
\end{array}\right) = \left[\mathds{1}-\mathcal{A}_0\right]^{-1}
\begin{pmatrix}
\mathds{G}_{0}^{1k}\\
\vdots\\
\mathds{G}_{0}^{(k - 1)k}\\
0\\
\mathds{G}_{0}^{(k + 1)k}\\
\vdots\\
\mathds{G}_{0}^{Nk}\end{pmatrix}
\label{Eq:MBGreen}
\end{equation}
for $k=1,...,N$ with 
\begin{equation}
\mathcal{A}_0 = \frac{\omega^2}{c^2}\left(\begin{array}{cccc}
0 & \mathds{G\mathrm{_{0}^{12}}}\alpha_{2} & \cdots & \mathds{G}_{\mathrm{0}}^{\mathrm{1N}}\mathrm{\alpha_{\mathrm{N}}}\\
\mathds{G}_{0}^{21}\alpha_{1} & \ddots & \ddots & \vdots\\
\vdots & \ddots & \ddots & \mathds{G}_{0}^{(N-1)N}\alpha_{N}\\
\mathds{G}_{0}^{N1}\alpha_{1} & \cdots & \mathds{G}_{0}^{N(N-1)}\alpha_{N-1} & 0\end{array}\right).
\label{Eq:A0}
\end{equation}
 With these relations and Eq.~(\ref{Eq:InterpartHeatFlux}) at hand
it is possible to determine the interparticle HF in a system
of $N$ particles out of equillibrium.

\begin{figure}[Hhbt]
\includegraphics[scale=0.6]{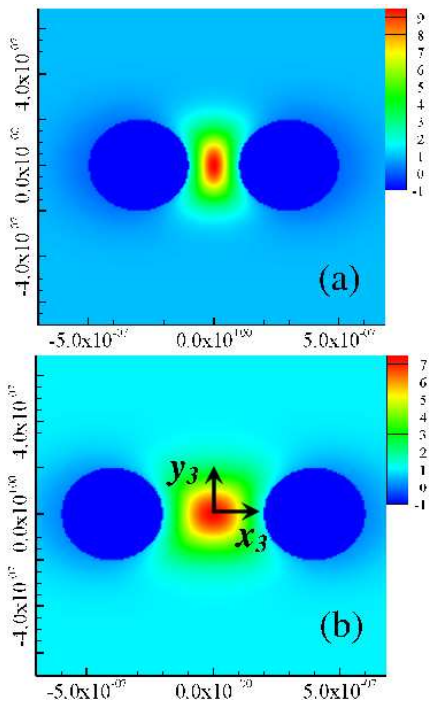}
\caption{Normalized HF exchanged between two SiC spherical particles maintained
at $T_{1}=300K$ (left particle) and $T_{2}=0\,{\rm K}$ (right particle)
with respect to the position $r_{3}=(x_{3},y_{3})$ of a third SiC
particle of same radius and $T_{3}=0\,{\rm K}$ for (a) $2 l = 600\,{\rm nm}$,
and (b) $2 l = 800\,{\rm nm}$. The HF is normalized by the HF exchanged
between two isolated dipoles in the same thermal conditions. The dark
zone with a negative HF corresponds to the region which cannot be
occupied by the third particle. For the sake of clarity we consider
here that all particles are identical (100 nm radius) and their electric
polarizability given by the simple Clausius-Mossotti form~\cite{Albaladejo} $\alpha=4\pi R^{3}\frac{\epsilon-1}{\epsilon+2}$, $R$
denoting the particles radius . The dielectric permittivity of particle
is described by a Drude-Lorentz model.} 
\end{figure}

\begin{figure}[Hhbt]
 \includegraphics[scale=0.4]{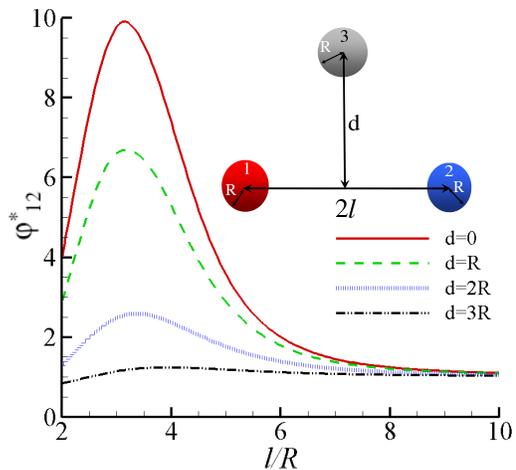}\caption{Normalized HF between two spherical particles of same radius
with respect to the position of a third one which is equidistant to
both particles ($T_{1}=300K$ , $T_{2}=T_{3}=0K$ ).}
\end{figure}

Let us now apply this theoretical formalism to describe some emerging
many-body effects. To this end, we consider the simplest possible
configuration where such effects occur that is a triplet
of particles. We consider only the interparticle HF between particle
1 and 2 separated by a distance $2l$ in the presence of the third
particle. Here, we assume that $T_{1}=300\,{\rm K}$ and $T_{2}=T_{3}=0$.
The interparticle HF is then given by 
\begin{equation}
  \varphi_{12}(2l,r_{3})=\mathcal{P}_{1\rightarrow2}-\mathcal{P}_{2\rightarrow1}=\mathcal{P}_{1\rightarrow2}.
\end{equation}
 In this case, the dyadic GF reads 
\begin{equation}
  \mathds{G}^{21}=\mathds{D}_{213}^{-1}\biggl[\mathds{G}_{0}^{21}+\mathds{B}^{213}\frac{\omega^{2}}{c^{2}}\mathds{D}_{31}^{-1}\mathds{G}_{0}^{31}\biggr]
\end{equation}
 with $\mathds{D}_{213}=\mathds{D}_{21}-\frac{\omega^{4}}{c^{4}}\mathds{B}^{213}\mathds{D}_{31}^{-1}\mathds{B}^{312}$,
$\mathds{D}_{21}=\mathds{1}-\frac{\omega^{4}}{c^{4}}\mathds{G}_{0}^{21}\alpha_{1}\mathds{G}_{0}^{12}\alpha_{2}$,
and $\mathds{B}^{213}=\mathds{G}_{0}^{23}\alpha_{3}+\frac{\omega^{2}}{c^{2}}\mathds{G}_{0}^{21}\alpha_{1}\mathds{G}_{0}^{13}\alpha_{3}$.
Some numerical results are shown in Fig.~1 and Fig.~2. We plot the
resulting interparticle HF $\phi_{12}^*$ between particle $1$ and $2$ in the
presence of body $3$ normalized to the HF for two isolated dipoles.
In both Figs.\ the position of the particles for which the interparticle
HF is calculated is fixed, but the position of the third particle
is changed. It can be seen that for some geometric configurations
the HF mediated by the presence of the third particle can be
larger than the value we usually measure for two isolated dipoles.
In particular, we observe an exaltation of HF of about one
order of magnitude when the third particle is located between the
two other particles, i.e., when all three particles are aligned. Hence,
the HF between two dipoles can dramatically be increased when
inserting a third particle in between. Note, that the three body system 
described above represent a photon heat transistor where the heat 
HF between two particles can be actively controlled  by the 
presence of a third particle. This could be achieved through 
classical AFM manipulation techniques.

\begin{figure}[Hhbt]
\includegraphics[scale=0.45]{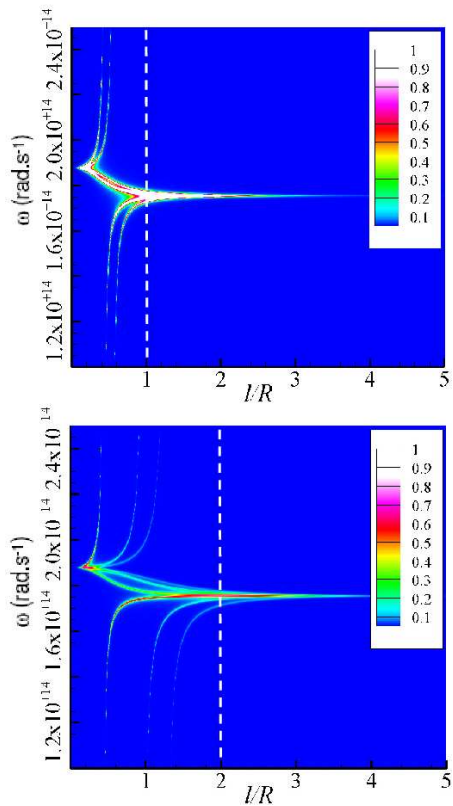}
\caption{TC between two spherical SiC particles separated
by a distance $2l$ (top) and between the same particles when a third
SiC particle is located on the mass center of this couple (bottom).
All particles have the same radius. The vertical dashed lines marks
the distance where the particles are touching~\cite{FOOTNOTE}.}
\end{figure}

In order to understand some of the physics behind this enhancement
mechanism, we show in Fig.~3 the TC $\mathcal{T}_{2,1}(\omega)$
for two isolated particles and three aligned particles for a frequency
range around the surface phonon resonance of the particles and for
different interparticle distances. First of all, it can be observed
that the TC shows different resonances where
it is close to $1$. These resonances can be found from the expression 
for the GF of the N-body system in Eq.~(\ref{Eq:MBGreen}) by 
evaluating $\det(\mathds{1} - \mathcal{A}_0) = 0$ where $\mathcal{A}_0$ is defined
in Eq.~(\ref{Eq:A0}). 
In fact, for the considered systems of two or three aligned particles one yields three configurational 
resonances~\cite{KellerEtAl1993}.  Two of these resonances are degenerate because of the rotational invariance around
the alignement axis (for more details see Ref.~\cite{SupplMat}). 
Apart from these configurational
resonances we have the surface mode resonance at $\omega=\omega_{sr}$
with $\epsilon(\omega_{sr})=-2$, which becomes dominant for large
distances so that one sees only one resonance in this case.
On the other hand, for distances close to the particle radius the
multiple interactions become dominant and several resonances show
up in the TC. Note, that the dipole 
model~\cite{Nara2008,Domingues2005} is only valid for $l>2R$. 
Now, from Fig.~3 it is clear that at long separation distances
the coupling between two dipoles becomes more efficient in presence
of a third mediator than without, so that the HF enhancement
can be attributed to a three body effect that is a resonant surface
mode coupling mediated by the third particle. Nevertheless, the absolute 
value for the interparticle HF is still far away from the theoretical 
upper limit.

In conclusion, we have introduced a theoretical framework to investigate
photon heat transport in mesoscopic systems where strong electromagnetic
interactions exist. In particular, a Meir-Wingreen-Landauer-type formula
for the radiative HF through N-body interacting photon regions
has been derived. A detailed study of three body systems has allowed
to identify a many body exaltation mechanism of HF due to configurational
resonances. This effect could  be used to improve for example
the performance of near-field thermophotovoltaic devices~\cite{MatteoEtAl2001,NarayanaswamyChen2003},
by placing nanoparticles on the surface of photovoltaic cells.

%
%

\begin{acknowledgments}
S.-A.\ B. gratefully acknowledges support from the Deutsche Akademie der Naturforscher Leopoldina
(Grant No.\ LPDS 2009-7). P. B. A. and K. J. acknowledge the
support of the Agence Nationale de la Recherche through the
Source-TPV project ANR 2010 BLANC 0928 01.
\end{acknowledgments}


\begin{thebibliography}{34}
\bibitem{FOOTNOTE} Here we also show the TC for
distances which cannot be achieved in a real situation because in that distance regime one can
easily distinguish the different resonances.
\end{thebibliography}
\end{document}